\magnification=\magstephalf
\input amstex
\loadbold
\documentstyle{amsppt}
\refstyle{A}
\NoBlackBoxes

\vsize=7.5in

\def\pf{\hfill $\square$}
\def\c{\cite}
 
\def\end{\text{End}}

\def\cd{\overset{\text{d}}\to\longrightarrow}
\def\cp{\overset{\text{P}}\to\longrightarrow }

\topmatter
\title The Beta-Hermite and Beta-Laguerre processes
\endtitle
\leftheadtext{L.-C. Li}
\rightheadtext{Beta-Hermite and Beta-Laguerre processes}

\author Luen-Chau Li\endauthor
\address{L.-C. Li, Department of Mathematics,Pennsylvania State University,
University Park, PA  16802, USA}\endaddress
\email luenli\@math.psu.edu\endemail

\abstract  In this work, we introduce matrix-valued diffusion
processes which describe the non-equilibrium situation of 
the matrix models for the beta-Hermite and the beta-Laguerre
ensembles. We also study the corresponding spectral measure 
process and empirical eigenvalue/singular value process 
with regard to their limit laws.

\endabstract
\endtopmatter

\document
\subhead
1. \ Introduction
\endsubhead

\baselineskip 15pt
\bigskip

The beta-ensembles have the physical interpretation as ensembles
of one dimensional Coulomb gas in a neutralizing background
with appropriate charge density \c{Dy,F}, where $\beta$ is 
the inverse temperature.  For $\beta =1, 2, 4,$ matrix models
of the beta-ensembles existed for a long time and are known
as the $*$-orthogonal, $*$-unitary, and $*$-symplectic ensemble respectively,
where $*\in \{\,\hbox{Gaussian, Laguerre, Jacobi}\,\}.$ (See for
example, \c{Meh} and \c{Muir} and the references therein.)   In
recent years, as a result of the work of Dumitriu and Edelman
\c{DE}, and that of Killip and Nenciu \c{KN}, matrix models
for the beta-ensembles with entries/parameters from classical 
distributions are now known for all values of $\beta >0.$   

In this work, we will restrict ourselves to considerations which are
related to the matrix models of the $\beta$-Hermite and the $\beta$-Laguerre
(Wishart) ensembles.  In order to explain
what we want to do, let us consider the matrix model of the $\beta$-Hermite
ensembles \c{DE}, defined schematically by
$$J_{\beta}\sim \pmatrix
\frac{1}{\sqrt{\beta}} N(0,1) & \frac{1}{\sqrt{2\beta}}\chi_{(n-1)\beta}  & 0 & 
\cdots\cr
\frac{1}{\sqrt{2\beta}}\chi_{(n-1)\beta} & \frac{1}{\sqrt{\beta}} N(0,1) & 
 \frac{1}{\sqrt{2\beta}}\chi_{(n-2)\beta} & \ddots\cr
0  & \frac{1}{\sqrt{2\beta}}\chi_{(n-2)\beta} & \frac{1}{\sqrt{\beta}} N(0,1) & 
\ddots\cr
\vdots & \ddots & \ddots & \ddots\cr
\endpmatrix ,\eqno(1.1)$$
where the entries on the diagonal and the subdiagonal 
of the $n\times n$ matrix $J_{\beta}$ independent of one another.
From the definition in (1.1), the joint density of the independent entries of 
$J_{\beta}$ is given by
$$W_{h}(a,b) = c_{n\beta} \prod_{k=1}^{n} b_{n-k}^{k\beta -1}
\exp\left[-\frac{\beta}{2}\left(\sum_{i=1}^{n} a^{2}_{i} + 2\sum_{i=1}^{n-1} b^{2}_{i}
\right)\right],\eqno(1.2)$$
where $a_i = (J_{\beta})_{ii},$ $b_{i} = (J_{\beta})_{i,i+1},$ and 
$c_{n\beta}$ is a normalization constant whose value is given in (3.4).
In this work, one of our motivating questions is the following: Is there 
a natural Markov process on Jacobi matrices for which $W_h(a,b)$ is the 
stationary steady state density ?   Our approach to this question can
be described as follows.  Since
the entries on the diagonal and the subdiagonal of $J_{\beta}$  in (1.1) 
are independent random variables,
therefore, it is natural to seek a Markov process which is a product of 
statistically
independent Markov processes on the diagonal and the subdiagonal
of the matrix.   In this
way, the problem reduces to that of constructing one-dimensional
Markov processes with the required properties.
As it turns out,  for the class of probability
density functions $p(x)$ satisfying Pearson's equation \c{P} (with
support on an interval $I\subset \Bbb R$ with endpoints $x_1$ and $x_2,$
say),
the problem of constructing a class of stationary Markov processes
with transition probability density $p(t, x_0, x)$ for which
$$\lim_{t\to\infty} p(t,x_0,x) = \int_{x_1}^{x_2} p(x_0)p(t,x_0,x)dx_0 =
p(x),\eqno(1.3)$$
has been the subject of investigation in \c{W}.  The Pearson's
family of probability density functions is quite broad, indeed it
includes many of the continuous distributions which we commonly
use such as normal, Chi-square, beta, etc.   For us, although
the Chi distribution is not on the list, however, the solution of
the problem for Chi-square suffices.  This is because the
Markov process corresponding to the pdf
of the Chi-square distribution can be identified to
be the square of a generalized Bessel process.   Hence
the correponding process for the Chi distribution must
be the generalized Bessel process itself and indeed,
the transition probability density function for this process
also satisfy the requirement in (1.3) above. (See Section 2.1 below.)
As for the normal distribution with mean $0,$ it is quite well-known 
that the Ornstein-Uhlenbeck
(OU) process describes the non-equilibrium situation \c{N}.  In fact, one
can readily understand Dyson's introduction of the matrix-valued 
Ornstein-Uhlenbeck processes in \c{Dy} from the perspective which we 
discussed above.
Once the matrix-valued diffusion processes are in place, there
are of course many questions one can ask about the
statistics of the corresponding eigenvalue processes.   In this work, we will
begin with basic results on the spectral measure and empirical
eigenvalue processes. (In the case of the $\beta$-Laguerre process,
we will instead deal with the empirical singular value process.)
Thus in a sense, we are trying to understand
the non-equilibrium properties of the Coulomb gas for general
values of $\beta >0$ and the these processes are basic.
In this connection, we remark that for $\beta =1,2, 4,$
the eigenvalue processes induced from the corresponding $\beta$-Hermite
diffusion processes are actually different from those in \c{Dy},
although there appears to be some subtle connections.  We
hope to return to this and other aspects of these processes
in subsequent publications.

The paper is organized as follows.  In Section 2, for the convenience
of the reader, we begin with a review of the generalized Bessel
and other related processes that play a key role in this work.
In Section 3, after a short description on the product of independent 
Markov processes,
we introduce the $\beta$-Hermite processes $(J_{\beta}(t))_{t\geq 0}$
for general values of $\beta >0$ in the first subsection. Then we 
calculate the distribution 
of the eigenvalues and the first components of the normalized
eigenvectors of $J_{\beta}(t)$ for each $t>0.$  The latter
quantity, as it turns out, is independent of $t.$
For the spectral measure $\mu_t =\sum_{j=1}^{n} \mu_j(t)\delta_{\lambda_j(t)}$
associated with $J_{\beta}(t),$
we also give the distribution of $\sum_{j=1}^{k} \mu_j(t)$ for each
$1\leq k\leq n.$  
In the second subsection, we consider the  
scaled process $(J^{(n)}_{\beta}(t))_{t\geq 0},$ where 
$J^{(n)}_{\beta}(t)= {J_{\beta}(t)\over \sqrt{n}}.$
The goal is to investigate the weak convergence (as $n\to\infty$),
in probability, of the corresponding spectral measure
$\mu^n_t$ and the empirical eigenvalue distribution $\nu^n_t$
for each $t>0.$   Because of the well-known one-to-one correspondence
between Jacobi matrices and spectral measures (see, for example,
\c{D1}), it is natural to consider $\mu^n_t.$   In this
connection, we make use of the method of moments to show
that $\mu^n_t$ converges weakly, in probability, to a
deterministic measure $\mu_t$ whose density function
has the shape of a semicircle.  In this way, we obtain a
time-dependent semicircle law.  Because the distribution of
the first components of the normalized eigenvectors of
$J_{\beta}(t)$ is independent of $t$  and the distribution
of $\sum_{j=1}^{k} \mu_j(t)$ is in fact identical to the one in
\c{BNR}, we can make use of the proof of Theorem 5.4 in \c{BNR}
to conclude that $d_{LP}(\mu^n_t,\nu^n_t)$ converges in
probability to $0$ as $n$ tends to infinity. (Here $d_{LP}$ is
the L\'evy-Prohorov metric.)  As the authors are studying
a rather different problem in \c{BNR}, it is quite remarkable
that we can make use of their analysis here.
In Section 4, we do the same for the $\beta$-Laguerre processes
and the associated $\beta$-Wishart processes.   In the case
of the scaled $\beta$-Wishart processes, the limiting law of both
the spectral measure and the empirical eigenvalue distribution
is that of a time-dependent Marchenko-Pastur law with 
a hard edge at $0$ and a soft upper edge at $4\rho(t)$, where
$\rho(t) = 1-e^{-t}.$  Finally,for the $\beta$-Laguerre processes, 
we show that the limit law for the empirical singular value
distribution is that of a time-dependent quarter circle law.

\bigskip
\bigskip

\subhead
2. \ The generalized Bessel processes
\endsubhead

\bigskip

Let $(b(t))_{t\geq 0}$ be standard Brownian motion on $\Bbb R,$
and $\delta >0.$  The Bessel process $(\tilde{R}^{\delta}(t))_{t\geq 0}$
of dimension $\delta$ starting from  $0$ \c{RY} is the 
square root of the process defined by the SDE
$$du(t) = 2 \sqrt{|u(t)|}\,db(t) + \delta\,dt,\,\,\,u(0)=0,\eqno(2.1)$$
where the point $0$ is an instantaneous reflecting barrier
for $0 <\delta <2$ and is polar otherwise.
When $\delta$ is a positive integer, it is well-known that 
$\tilde{R}^{\delta}(t)$ is the 
modulus of a standard Brownian motion in $\Bbb R^{\delta}.$  
For our purpose, recall that the the transition
probability density function of $\tilde{R}^{\delta}(t)$ is given by
$$\tilde{p}_{t}^{\delta}(x_0,x) = 
t^{-1}\left(\frac{x}{x_0}\right)^{\frac{\delta}{2} -1} x
\exp\left(-\frac{x_0^2 + x^2}{2t}\right)
\,I_{\frac{\delta}{2}-1}\left(\frac{xx_0}{t}\right) \,\,\,\hbox{for}\,x_0>0,
\eqno(2.2)$$
and
$$\tilde{p}_{t}^{\delta}(0,x) = 2^{-\frac{\delta}{2} +1}t^{-\frac{\delta}{2}}\,
\Gamma\left(\frac{\delta}{2}\right)^{-1}
  x^{\delta -1} \exp\left(-\frac{x^2}{2t}\right),\eqno(2.3)$$
where $\Gamma(x)$ is the Gamma function, and  $I_{\nu}$ is the modified 
Bessel function of order $\nu.$  Since we have $\tilde{R}^{\delta}(0)=0$
with our definition, it follows that the pdf $\tilde{p}_{t}^{\delta}(x)$
of $\tilde{R}^{\delta}(t)$ coincides with $\tilde{p}_{t}^{\delta}(0,x).$
We will also deal with the Ornstein-Uhlenbeck processes in this work.  By
definition, the one dimensional Ornstein-Uhlenbeck process
with parameters $(a, \sigma)$ starting from $0$ is the solution
of  the SDE
$$dv(t) = -a v(t)\,dt + \sigma\,db(t),\,\,\, v(0)=0,\eqno(2.4)$$
where $a > 0$ and $\sigma\in \Bbb R\setminus\{0\}.$  By the 
Dambis-Dubins-Schwarz theorem \c{RY},
$v(t)$ is a time-changed Brownian motion:
$$v(t) = \sigma e^{-at} w\left(\frac{e^{2at}-1}{2a}\right),\eqno(2.5)$$
where $w(t)$ is a standard Brownian motion in $\Bbb R.$  Consequently,
the random variables $v(t)$ are Gaussian with mean zero and 
variance
$$\rho(t) = \frac{\sigma^2}{2a} (1-e^{-2at}),\eqno(2.6)$$
and the transition probability density function is given by
$$\widehat{p}_{t}(x_0,x) = \frac{1}{\sqrt{2\pi\rho(t)}}\exp\left(
-\frac{(x - x_0 e^{-at})^2}{2\rho(t)}\right).\eqno(2.7)$$

Note that if \, $\bold{v}(t) = (v_1(t),\cdots,v_d(t))$, where 
$v_1(t),\cdots, v_d(t)$
are independent one dimensional Ornstein-Uhlenbeck processes starting
from $0$ with
the same parameters $(a,\sigma)$, then from (2.5) above, we have
$$v_i(t) = \sigma e^{-at} w_i\left(\frac{e^{2at}-1}{2a}\right),\,\,\,i=1,\cdots,
d,\eqno(2.8)$$
where $w_1(t),\cdots,w_d(t)$ are independent standard Brownian motions
on $\Bbb R.$ Thus it follows from (2.8) that the radial part of the
$d$-dimensional Ornstein-Uhlenbeck process $\bold{v}(t)$ is given by
$$\eqalign{
R^{d}(t):&= \sqrt{(v_1(t))^2 + \cdots + (v_d(t))^2}\cr
        &=\sigma e^{-at}\tilde{R}^{d}\left(\frac{e^{2at}-1}{2a}\right),\cr}
\eqno(2.9)$$
where $\tilde{R}^{d}$ is a Bessel process of dimension $d$ starting from
$0.$  Thus this motivates the following definition.
\smallskip

\definition
{Definition 2.2} Let $\delta >0,$ and let $a$ and $\sigma$ be as
above.  If $(\tilde{R}^{\delta}(t))_{t\geq 0}$ is a Bessel process
of dimension $\delta$ starting from $0,$ then the process 
$(R^{\delta}(t))_{t\geq 0}$
defined by
$$R^{\delta}(t)
  =\sigma e^{-at}\tilde{R}^{\delta}\left(\frac{e^{2at}-1}{2a}\right)\eqno(2.10)$$
is called the generalized Bessel process of dimension $\delta$ starting
from $0$ with parameters $(a,\sigma).$
\enddefinition

\smallskip

From the above definition, a straightforward calculation
using (2.2) and (2.3) shows that the pdf and the transition
probability density function of $R^{\delta}(t)$ are given by
$$p^{\delta}_t(x)= p^{\delta}_{t}(0,x)= 
2^{-\frac{\delta}{2} +1}\rho(t)^{-\frac{\delta}{2}}\,
\Gamma\left(\frac{\delta}{2}\right)^{-1}
  x^{\delta -1} \exp\left(-\frac{x^2}{2\rho(t)}\right)\,1_{[0,\infty)}(x)
\eqno(2.11)$$
and
$$\eqalign{
& p^{\delta}_{t}(x_0,x) \cr
= & \rho(t)^{-1}\left(\frac{x}{e^{-at}x_0}\right)^{\frac{\delta}{2} -1} x
\exp\left[-\frac{(x^2 + e^{-2at}x_0^2)}{2\rho(t)}\right]
I_{\frac{\delta}{2}-1}\left(\frac{xe^{-at}x_0}{\rho(t)}\right),\,x_0>0.\cr}
\eqno(2.12)$$
\smallskip

\remark
{Remark \rom{2.3}} In \c{Eie}, the author used the term
generalized Bessel process to denote the radial part of a $d$-dimensional 
Ornstein-Uhlenbeck process.
Here we are adapting this terminology to the more general
situation when $\delta$ is not necessarily an integer.
\endremark
\smallskip
If $\widehat{p}_{t}(x)$ denote the pdf of the Ornstein-Uhlenbeck process
above, it is well-known that \c{N}
$$\lim_{t\to\infty} \widehat{p}_{t}(x_0,x) = \int_{-\infty}^{\infty} 
\widehat{p}_{\infty}(x_0)\widehat{p}_{t}x_0,x)dx_0 = 
\widehat{p}_{\infty}(x),\eqno(2.13)$$
where $\widehat{p}_{\infty}(x)$ is the pdf of the normal
distribution $N(0,\rho(\infty)).$

\proclaim
{Proposition 2.4} For the transition probability density function
of the generalized Bessel process in (2.11),(2.12),
$$\lim_{t\to\infty} p^{\delta}_{t}(x_0,x) = \int_{0}^{\infty} 
p^{\delta}_{\infty}(x_0)p^{\delta}_{t}(x_0,x)dx_0 = p^{\delta}_{\infty}(x),\eqno(2.14)$$
where
$$p^{\delta}_{\infty}(x)= 2^{-\frac{\delta}{2} + 1} 
\rho(\infty)^{-\frac{\delta}{2} + 1}
\Gamma\left(\frac{\delta}{2}
\right)^{-1} \,x^{\delta -1}\exp\left(-\frac{x^2}{2\rho(\infty)}\right).
\eqno(2.15)
$$
\endproclaim

\demo
{Proof} The transition probability density function $q^{\delta}_{t}(x_0,x)$
of the square of the generalized Bessel process above is related
to $p^{\delta}_{t}(x_0,x)$ by
$$q^{\delta}_{t}(x_0,x)=\frac{1}{2\sqrt{x}} p^{\delta}_{t}(\sqrt{x_0},\sqrt{x}).
\eqno(2.16)$$
Hence the explicit expression for $q^{\delta}_{t}(x_0,x)$ can 
be computed from (2.11) and (2.12).  Note that the function 
$q^{\delta}_{\infty}(x) = 2^{-1} x^{-1/2} p^{\delta}_{\infty}(\sqrt{x})$
(resp.~ $q^{\delta}_{t}(x_0,x)$)
is related to the one in Eqn.(28) of \c{W} (resp.~ Eqn.(30) of \c{W})
by
scaling of the time variable, the spatial variables and
an overall scaling of the function itself.  Hence
it follows from \c{W} that
$$\lim_{t\to\infty} q^{\delta}_{t}(x_0,x) = \int_{0}^{\infty} 
q^{\delta}_{\infty}(x_0)q^{\delta}_{t}x_0,x)dx_0 = q^{\delta}_{\infty}(x),
\eqno(2.17)$$
where 
$$q^{\delta}_{\infty}(x) = 
 2^{-\frac{\delta}{2}} \rho(\infty)^{-\frac{\delta}{2}}
\Gamma\left(\frac{\delta}{2}
\right)^{-1} \,x^{\frac{\delta}{2} -1}\exp\left(-\frac{x}{2\rho(\infty)}\right).
\eqno(2.18)$$
Since $p^{\delta}_{\infty}(x) = 2x\,q^{\delta}_{\infty}(x^2),$
it is immediate from (2.17) that
$\lim_{t\to\infty} p^{\delta}_{t}(x_0,x) = p^{\delta}_{\infty}(x).$  Moreover,
on using (2.15) and (2.17), we obtain
$$\eqalign{
 \int_{0}^{\infty} p^{\delta}_{\infty}(x_0)p^{\delta}_{t}(x_0,x)\,dx_0
=\, & 2x\int_{0}^{\infty} q^{\delta}_{\infty}(y_0)q^{\delta}_{t}(y_0,x^2)\,dy_0\cr
=\, & 2x q^{\delta}_{\infty}(x^2) = p^{\delta}_{\infty}(x).\cr}\eqno(2.19)$$
So this completes the proof.
\pf
\enddemo

\remark
{Remark \rom{2.5}} (a) Alternatively, one could give a direct proof of
the limiting behaviour of $p^{\delta}_{t}(x_0,x)$ for $x_0>0$ by invoking 
the asymptotics $I_{\nu}(z)\sim (z/2)^{\nu}/\Gamma(\nu+1)$
as $z\to0.$  On the other hand,  the fact that $p^{\delta}_{\infty}(x)$
is the stationary density of the generalized Bessel process 
can also be established in a direct way by making use 
of the Weber Sonine formula \c{Wat}
$$\eqalign{
& \int_{0}^{\infty} x^{\mu} e^{\alpha x^2} J_{\nu}(\gamma x)\,dx\cr
=\, &\frac{\beta^{\nu}\Gamma\left(\frac{1}{2}(\mu + \nu + 1)\right)}
{2^{\nu +1} \alpha^{\frac{1}{2} (\mu+\nu+1)}\Gamma(\nu+1)}
\,_1F_1\left(\frac{1}{2} (\mu+\nu+1), \nu+1; -\frac{\gamma^2}{4\alpha}\right),
\cr}\eqno(2.20)$$
valid for $\hbox{Re}\,\alpha >0$, $\hbox{Re}(\mu+\nu) >-1.$
We will leave the details to the interested reader.
\newline
(b) In \c{W}, the author applies the method of separation of
variables to the Fokker-Planck equation and identifies the
cofficients of this equation with those of the Pearson's
equation.  In this way, eigenvalue problems of Sturm-Liouville
type with reflecting boundary conditions at the end points
are obtained.   It is interesting to point out that as a
result of the connection which we mentioned above, we have
the expansion
$$\eqalign{&
p^{\delta}_{t}(x_0,x)\cr
=\, & p^{\delta}_{\infty}(x)\sum_{n=0}^{\infty} nB\left(n,\frac{\delta}{2}\right)
e^{-2ant}L^{\frac{\delta}{2}-1}_{n}\left(\frac{x_0^2}{2\rho(t)}\right)
L^{\frac{\delta}{2}-1}_{n}\left(\frac{x^2}{2\rho(t)}\right),\cr}
\eqno(2.21)
$$
where $B(a,b)$ is the Beta function and
$$L^{\alpha}_{n}(x) = \frac{1}{n!} x^{-\alpha} e^{-x} \frac{d^n}{dx^n}
(x^{\alpha} e^{-x})\eqno(2.22)$$
are the Laguerre polynomials.
\endremark
\smallskip
For our construction in the rest of the paper, we will pick the
normalization $(a,\sigma) =(1/2,1).$  In this case, $p^{\delta}_{\infty}(x)$
is just the pdf of the Chi distribution $\chi_{\delta}$
while $\widehat{p}_{\infty}(x)$ is 
that of standard normal.  Hence the corresponding generalized Bessel
process and Ornstein-Uhlenbeck process have the desired properties. 
(See (1.1) above.) 

\bigskip
\bigskip

\subhead
3. \ The beta-Hermite processes
\endsubhead

\bigskip

We begin by defining the product Markov process which will be used
in our construction in the present section and the next.
For this purpose, let $X=(X_{t})_{t\geq 0}$ and $Y=(Y_{t})_{t\geq 0}$ be 
two statistically
independent time-homogeneous Markov processes 
with transition probability density functions given by 
$p_{X}(t,x_0,x)$ and $p_{Y}(t,y_0,y)$ respectively.  Then the
product process $X\otimes Y$ is defined to be the process where 
$(X\otimes Y)_t = (X_t,Y_t)$  for $t\geq 0.$  Put
$p_{X\otimes Y}(t, (x_0,y_0), (x,y)) = p_{X}(t,x_0,x)p_{Y}(t,y_0,y),$
then it can be verified that the product process is again a Markov 
process with $p_{X\otimes Y}(t, (x_0,y_0), (x,y))$  as its transition
probability density function.  Clearly, we can iterate this
construction and so we can define the product Markov process
for any given number of statistically independent time-homogeneous
Markov processes.

\subhead
3.1 \ The beta-Hermite processes and their eigenvalue distribution
\endsubhead
\medskip

\definition
{Definition 3.1.1} The $\beta$-Hermite process $(J_{\beta}(t))_{t\geq 0}$
is the stochastic process on $n\times n$ Jacobi matrices $J_{\beta}(t)$
given by
$$J_{\beta}(t) = \pmatrix
\frac{1}{\sqrt{\beta}} U_1(t) & \frac{1}{\sqrt{2\beta}}R^{(n-1)\beta}(t)  & 0 & 
\cdots\cr
\frac{1}{\sqrt{2\beta}}R^{(n-1)\beta}(t) & \frac{1}{\sqrt{\beta}} U_2(t) & 
 \frac{1}{\sqrt{2\beta}}R^{(n-2)\beta}(t) & \ddots\cr
0  & \frac{1}{\sqrt{2\beta}}R^{(n-2)\beta}(t) & \frac{1}{\sqrt{\beta}} U_3(t) & 
\ddots\cr
\vdots & \ddots & \ddots & \ddots\cr
\endpmatrix\eqno(3.1)$$
where the processes on the diagonal and the subdiagonal are
statistically independent of each other.  Here, 
$U_1(t), \cdots, U_n(t)$ are  Ornstein-Uhlenbeck proceeses starting from
$0$ with parameters $(1/2, 1)$
and $R^{j\beta}(t)$ is the generalized Bessel process 
of dimension $j\beta$ starting from $0$ with parameters 
$(1/2, 1),\,\,j=1,\cdots,n-1.$
\enddefinition

\proclaim
{Proposition 3.1.2} The $\beta$-Hermite process is a matrix-valued
diffusion process starting from $0$ with transition probability
density function 
$$P(t,\widetilde{J},J)= 2^{\frac{n}{2}} \beta^{n-\frac{1}{2}}
  \prod_{i=1}^{n} \widehat{p}_{t}(\sqrt{\beta}\,\,\tilde{a}_i, \sqrt{\beta}\,\,a_i)
  \prod_{j=1}^{n-1} p^{(n-j)\beta}_{t} (\sqrt{2\beta}\,\,\tilde{b}_j,\sqrt{2\beta}\,\,
  b_j),
\eqno(3.2)$$
with respect to Lebesgue measure $dadb = da_1\cdots da_{n}db_1\cdots db_{n-1}$
on $\Bbb R^{n} \times \Bbb R_{+}^{n-1},$ where $\tilde{a}_i =\widetilde{J}_{ii}, 
\,\tilde{b}_{i}=\widetilde{J}_{i,i+1},$
$a_i = J_{ii}$ and $b_{i}= J_{i,i+1}.$  Consequently, the
joint density of the independent entries of $J_{\beta}(t)$ is given by
$$P(t,0,J) = c_{n\beta}\, \rho(t)^{-{n\over 2}-{\beta\over 4} n(n-1)}
\prod_{k=1}^{n-1} b_{n-k}^{k\beta-1} \exp\left(-\frac{\beta}{2\rho(t)} tr 
J^{2}\right),\eqno(3.3)
$$
where
$$c_{n\beta} =\frac{2^{{n\over 2}-1}\beta^{{n\over 2}+{\beta\over 4}n(n-1)}}
{\pi^{n\over 2}\prod_{k=1}^{n-1}\Gamma\left(k\beta\over 2\right)}.\eqno(3.4)$$
Finally, if $W_{h}(a,b)$ is the joint density of the independent entries of
the $\beta$-Hermite ensemble in (1.1), then
$$\lim_{t\to\infty} P(t,\widetilde{J}, J) = \int_{\Bbb{R}^n\times \Bbb{R}_{+}^n} 
W_{h}(\tilde{a},\tilde{b})P(t,\widetilde{J},J)\,\,d\tilde{a}d\tilde{b} = 
W_{h}(a,b).\eqno(3.5)$$ 

\endproclaim

\demo
{Proof} It is easy to see that the transition probability density functions
of $\frac{1}{\sqrt{2\beta}}R^{\delta}(t)$ and $\frac{1}{\sqrt{\beta}}v(t)$
are given by
$\sqrt{2\beta}\, p^{\delta}_{t}(\sqrt{2\beta}\,x_0,\sqrt{2\beta}\,x)$
and 
$\sqrt{\beta}\, \widehat{p}_{t}(\sqrt{\beta}\,x_0,\sqrt{\beta}\,x)$
respectively.  Therefore, (3.2) is a consequence of the product
construction of independent Markov processes.  As
$P[J_{\beta}(0) =0] =1,$ the pdf in (3.3) now follows
from (3.2), (2.11), and the explicit formula for 
$\widehat{p}_{t}(x).$  Finally, the validity of (3.5)
is due to (2.13) and (2.14) and this completes the verification.

\pf
\enddemo

Our next goal is to calculate the joint probability density
function of the eigenvalues of $J_{\beta}(t).$  In preparation, 
observe that the pdf of 
$\frac{1}{\sqrt{\beta}} U_i(t)$  
can be expressed in terms of the pdf $\widehat{p}_{\infty}(x)$ of 
$N(0,1).$  Indeed, this pdf is given by
$$\eqalign{\sqrt{\beta}\, \widehat{p}_{t}(\sqrt{\beta}\,x)= &
\left(\frac{\beta}{2\pi\rho(t)}\right)^{1/2}\exp\left(-\frac{\beta x^2}
{2\rho(t)}\right)\cr
= & \sqrt{\frac{\beta}{\rho(t)}} \,\,\widehat{p}_{\infty}\left(
\sqrt{\frac{\beta}{\rho(t)}}\,\, x\right).\cr}\eqno(3.6)$$
Similarly, it follows from (2.11) that the pdf of   
$\frac{1}{\sqrt{2\beta}}R^{j\beta}(t)$ takes the form
$$\eqalign{
\sqrt{2\beta}\, p^{j\beta}_{t}(\sqrt{2\beta}\,x) = &
\frac{2\left(\beta\over \rho(t)\right)^{j\beta/2}}{\Gamma\left(\frac{j\beta}{2}
\right)}\,x^{j\beta -1} \exp\left(-\frac{\beta x^2}{\rho(t)}\right)\,1_{[0,\infty)}(x)\cr
= & \sqrt{\frac{2\beta}{\rho(t)}} \,\,p^{j\beta}_{\infty}\left(
\sqrt{\frac{2\beta}{\rho(t)}}\,\, x\right),\cr}\eqno(3.7)$$ 
where $p^{j\beta}_{\infty}(x)$ is the pdf of the Chi distribution $\chi_{j\beta}.$
Hence we have
$$\eqalign{
P(t,0,J) = &\, 2^{\frac{n}{2}} \left(\frac{\beta}{\rho(t)}\right)^{n-\frac{1}{2}}
  \prod_{i=1}^{n} \widehat{p}_{\infty}(\sqrt{\beta/\rho(t)}\,\,a_i)
  \prod_{j=1}^{n-1} p^{(n-j)\beta}_{\infty} (\sqrt{2\beta/\rho(t)}\,\,b_j)\cr
  = &\, \frac{1}{\rho(t)^{n-\frac{1}{2}}} W_{h} \left(\frac{a}{\sqrt{\rho(t)}}, 
      \frac{b}{\sqrt{\rho(t)}}\right).\cr}
\eqno(3.8)$$

Now recall that a generic Jacobi matrix $J=J(a,b)$ has nonzero entries on 
the subdiagonal and the eigenvalues of $J$ are distinct.  We will
order the eigenvalues such that
$\lambda_{1}>\lambda_{2}>\cdots \lambda_{n}$ and denote 
by $f_{1}(1) >0,f_{2}(1)>0,\cdots, f_{n}(1)>0$
the first components of the normalized eigenvectors corresponding to
the distinct eigenvalues.  Put 
$\lambda =(\lambda_1,\cdots,\lambda_n)$, $f(1) = (f_1(1),\cdots, f_n(1)).$
It is well-known that the map
$\phi: J(a,b)\longrightarrow (\lambda, f(1))$ is a 
diffeomorphism from the set of generic Jacobi matrices with positive
subdiagonal entries to $C_{+}\times S^{n-1}_{+},$ where 
$C_{+}=\{(x_1,\cdots, x_n)\in \Bbb R^n\mid x_1 >\cdots >x_n \},$
and $S^{n-1}_{+}$ consists of vectors $q=(q_1,\cdots,q_n)$ on the unit sphere
$S^{n-1}$ such that $q_i >0$ for all $i$ (see, for example, \c{D1}).
Moreover, it follows from  \c{DE},\c{D2} that
$$da\,db = \frac{\prod_{i=1}^{n-1} b_i}{\prod_{i=1}^{n} f_{i}(1)}d\lambda\,d\sigma ,
\eqno(3.9)$$
where
$$d\sigma = \frac{df_{1}(1)\cdots df_{n-1}(1)}{f_{n}(1)}\eqno(3.10)$$ 
is the element of surface area in $S^{n-1}_{+} .$
\proclaim
{Theorem 3.1.3} Under the $\beta$-Hermite process, the vector 
$\lambda(t)=(\lambda_1(t),\cdots,\lambda_n(t))$ of 
ordered eigenvalues of $J_{\beta}(t)$ and the vector 
$f(1,t) =(f_1(1,t),\cdots, f_n(1,t))$ of
first components of normalized eigenvectors are independent.
If $\Delta (\lambda)$ is the Vandermonde determinant,
then 
\smallskip
\noindent (a) the joint pdf of the unordered eigenvalues of  
$J_{\beta}(t)$ is given by
$$\eqalign{
P_{n\beta}(t,\lambda) & = C_{n\beta}\, \rho(t)^{-{n\over 2}-{\beta\over 4} n(n-1)}
|\Delta(\lambda)|^{\beta} \exp \left(-\frac{\beta}{2\rho(t)} \sum_{i=1}^{n}
\lambda_{i}^{2}\right)\cr
& = C_{n\beta}\, \rho(t)^{-{n\over 2}-{\beta\over 4} n(n-1)} \exp (-\beta W(t,\lambda)),
\cr}
\eqno(3.11)$$
where 
$$C_{n\beta}= (2\pi)^{-{n\over 2}} \beta^{{n\over 2}+{\beta\over 4}n(n-1)}\prod_{j=1}^{n}
\frac{\Gamma\left(1 + {\beta\over 2}\right)}
{\Gamma\left(1 + {j\beta\over 2}\right)},\eqno(3.12)$$
and
$$W(t,\lambda) = \frac{1}{2\rho(t)} \sum_{i=1}^{n} \lambda_i^2 - \sum_{i<j} 
\log|\lambda_i-\lambda_j|.\eqno(3.13)$$
\smallskip
\noindent (b) the joint density of $f_1(1,t),\cdots, f_n(1,t)$ with respect to
the measure $d\sigma$ on $S^{n-1}_{+}$ is given by
$$2^{n-1} \frac{\Gamma\left(n\beta\over 2\right)}{\left(\Gamma\left(\beta\over 2
\right)\right)^{n}} \prod_{i=1}^{n} q^{\beta -1}_{i}.\eqno(3.14)$$
\endproclaim

\demo
{Proof} Let $\phi$ be the map above and consider $J_{\beta}(t)$ in
the domain of $\phi.$  Let $\lambda_1(t) >\cdots >\lambda_n(t)$ be
the (ordered) eigenvalues of $J_{\beta}(t)$ and let 
$f_1(1,t)>0,\cdots f_n(1,t)>0$ be the first components of
the normalized eigenvectors corresponding to the eigenvalues.
Because of (3.8), the independence of $\lambda(t)$ and
$f(1,t)$ follows as in the calculation in \c{DE}
where we have to use (3.9) and the relation \c{D1},\c{DE}
$$\Delta(\lambda) \equiv \prod_{i<j} (\lambda_i-\lambda_j)
  = \frac{\prod_{i=1}^{n-1} b^{i}_{n-i}}{\prod_{i=1}^{n} f_{i}(1)}.\eqno(3.15)$$
The calculations leading to the assertions in the remaining parts
of the proposition are also similar and so we skip the details.
\pf
\enddemo

Now, another way to describe generic Jacobi matrices $J(a,b)$ is by
using the spectral measure 
$$\mu = \sum_{j=1}^{n} \mu_j\,\delta_{\lambda_j},\quad \mu_j = f_j(1)^2,\,\,\,
1\leq j \leq n.\eqno(3.16)$$
Indeed, it is well-known that the map $\psi: J(a,b)\longrightarrow \mu$
is a bijection from the set of generic $n\times n$ Jacobi matrices 
to the set of probability measures on $\Bbb R$ supported at $n$ points.
Let
$$\mu(t) = (\mu_1(t),\cdots, \mu_n(t)),\quad \mu_j(t) =f_j(1,t)^2,\,\,1\leq j \leq n.\eqno(3.17)$$
To describe the probability distribution of the vector $\mu(t),$ recall
that the Dirichlet distribution $\hbox{Dir}_{n-1}(\alpha_1,\cdots; \alpha_n)$
with parameters $\alpha_1,\cdots,\alpha_n >0$ is the distribution which
has a density with respect to Lebesgue measure on $\Bbb R^{n-1}$ given by
(see \c{Wi} for more details)
$$\frac{\Gamma(\alpha_1 + \cdots + \alpha_n)}{\Gamma(\alpha_1)\cdots\Gamma(\alpha_n)}\prod_{j=1}^{n} y_j^{\alpha_j-1}\,1_{S}(y_1,\cdots,y_{n-1})\eqno(3.18)$$
where
$$y_n =1-y_1-\cdots -y_{n-1},\eqno(3.19)$$
and where $S$ is the simplex
$$S=\left\{(y_1,\cdots,y_{n-1})\Big|\,\, y_j\geq 0\,\, \hbox{for all}\,\,j, 
\sum_{j=1}^{n-1}y_j\leq 1\right\}.\eqno(3.20)$$ 
Note that for $n=2,$ $\hbox{Dir}_{1}(\alpha_1;\alpha_2)$ is the
Beta distribution $\hbox{Beta}(\alpha_1,\alpha_2)$ which is
supported on $[0,1].$  The following result is a straightforward
consequence of the joint density of $f_1(1,t),\cdots, f_n(1,t)$
with respect to $d\sigma$ and the basic
properties of the Dirichlet distribution \c{Wi}.

\proclaim
{Corollary 3.1.4} Under the $\beta$-Hermite process, the vector
of weights $\mu(t)$ of the spectral measure associated with
$J_{\beta}(t)$ follows the
distribution ${\hbox{Dir}}_{n-1}({\beta\over 2},\cdots; {\beta\over 2}).$
Hence the marginals are Beta distributions:
$$\mu_j(t)\sim \hbox{Beta}\left({\beta\over 2}, {(n-1)\beta\over 2}\right), 
\,\,1\leq j\leq n.\eqno(3.21)$$
Moreover, for each $1\leq k\leq n,$
$$\sum_{j=1}^{k} \mu_{j}(t) \sim \hbox{Beta} \left({k\beta\over 2}, 
{(n-k)\beta\over 2}\right).\eqno (3.22)$$
\endproclaim

\subhead
3.2 \ Time-dependent semicircle law
\endsubhead
\medskip

We introduce the following scaling of $J_{\beta}(t)$:
$$J^{(n)}_{\beta}(t) = \frac{J_{\beta}(t)}{\sqrt{n}},\eqno(3.23)$$
and let $a^{(n)}_{k}(t) = (J^{(n)}_{\beta}(t))_{kk}$, 
$b^{(n)}_{k}(t) = (J^{(n)}_{\beta}(t))_{k,k+1}.$  
The goal of this subsection is to study the large $n$ behaviour of
the spectral measure process 
$$\mu^{n}_{t} = \sum_{j=1}^{n} \mu^{(n)}_{j}(t) \delta_{\lambda_{j}^{(n)}(t)}\eqno(3.24)$$
and the empirical eigenvalue process
$$\nu^{n}_{t} = {1\over n} \sum_{i=1}^{n} \delta_{\lambda^{(n)}_i(t)}\eqno(3.25)$$
for each $t>0,$ where $\lambda^{(n)}_{1}(t),\cdots, \lambda^{(n)}_{n}(t)$ are the 
eigenvalues of $J^{(n)}_{\beta}(t).$   We begin with a lemma.

\proclaim
{Lemma 3.2.1} As $n\to\infty,$
$$a^{(n)}_{k}(t)\cp 0,\quad b^{(n)}_{k}(t)\cp \sqrt{\rho(t)\over 2}\eqno(3.26)$$
for each $t>0.$
\endproclaim
\demo
{Proof}  For any $t>0$ and any $r =1,2,\cdots,$ it follows from (3.6) that
$$\eqalign{
E[(a^{(n)}_k(t))^r] & = \int_{-\infty}^{\infty}\left(\frac{n\beta}{2\pi\rho(t)}\right)^{1/2} x^{r}\exp\left(-\frac{n\beta x^2}{2\rho(t)}\right)\,dx\cr
& = {\frac{1}{\sqrt{2\pi}}}\left({\frac{\rho(t)}{n\beta}}\right)^{r\over 2}
\int_{-\infty}^{\infty} x^{r} e^{-{x^{2}\over 2}}\, dx \cr}\eqno(3.27)$$
from which it is clear that $E[(a^{(n)}_k(t))^r]\to 0$ as $n\to\infty.$
Therefore, $a^{(n)}_k(t)\cd 0$ and hence $a^{(n)}_k(t)\cp 0.$  Similarly,
we obtain from (3.7) that
$$\eqalign{
E[(b^{(n)}_k(t))^r] & = \frac{2\left(n\beta\over \rho(t)\right)^{(n-k)\beta/2}}
{\Gamma\left(\frac{(n-k)\beta}{2}\right)}\int_{0}^{\infty} 
\,x^{(n-k)\beta -1+r} \exp\left(-\frac{n\beta x^2}{\rho(t)}\right)\,dx\cr
& = \frac{2\left(\frac{\rho(t)}{n\beta}\right)^{r\over 2}}
{\Gamma\left(\frac{(n-k)\beta}{2}\right)}\int_{0}^{\infty} 
x^{(n-k)\beta + r -1} e^{-x^2}\,dx\cr
& = \left(\frac{\rho(t)}{n\beta}\right)^{r\over 2}
    \frac{\Gamma\left(\frac{(n-k)\beta+r}{2}\right)}
         {\Gamma\left(\frac{(n-k)\beta}{2}\right)}.\cr}\eqno(3.28)$$
From the asymptotics of the Gamma function, we have
$$ \frac{\Gamma\left(\frac{(n-k)\beta+r}{2}\right)}
         {\Gamma\left(\frac{(n-k)\beta}{2}\right)}\sim  
         \left({n\beta}\over 2\right)^{r\over 2}\eqno(3.29)$$
as $n\to\infty.$   Hence the assertion  $b^{n}_{k}(t)\cp \sqrt{\rho(t)\over 2}$
follows from (3.28) and (3.29).
\pf
\enddemo

We now turn to the analysis of the
spectral measure process $(\mu^{n}_{t})_{t\geq 0}.$   First, from the relations
$$\int_{\Bbb R} x^{k} d\mu^{n}_{t}(x) = (e_1, (J^{(n)}_{\beta}(t))^{k}\,e_1),\,\,k=1,\cdots,
\eqno(3.30)$$
it is clear that the moments $\int_{\Bbb R} x^{k} d\mu^{n}_{t}(x)$ are
polynomials in the entries of $J^{(n)}_{\beta}(t)$.  As the entries on
the diagonal and subdiagonal of
$J^{(n)}_{\beta}(t)$ are independent random variables, it follows from
(3.26) that as $n\to\infty,$
$$(a^{(n)}_{1}(t),b^{(n)}_{1}(t),\cdots, a^{(n)}_{j}(t),b^{(n)}_{j}(t))\cd
\left(0,\sqrt{\rho(t)\over 2}, \cdots, 0,\sqrt{\rho(t)\over 2}\right)
\eqno(3.31)$$
for each fixed value of $j$ and each $t>0.$   Hence by the continuous
mapping theorem, (3.30) and (3.31), we obtain
$$\int_{\Bbb R} x^{k} d\mu^{n}_{t}(x)\cp (e_1, (J^{(\infty))}(t))^{k}\,e_1)\eqno(3.32)$$
as $n\to\infty,$
where $J^{(\infty)}(t)$ is the Jacobi operator on $\ell_{2}^{+}$ given by
$$J^{(\infty))}(t)= \pmatrix 0 & \sqrt{\rho(t)\over 2} & 0 & 0 & \cdots  \\
               \sqrt{\rho(t)\over 2} & 0 &\sqrt{\rho(t)\over 2} & 0 &\cdots\\
               0 & \sqrt{\rho(t)\over 2}& 0 &\sqrt{\rho(t)\over 2}& \cdots \\
                \cdots & \cdots & \cdots & \cdots &\cdots \endpmatrix. 
\eqno(3.33)$$
Now the orthogonal polynomials corresponding to $J^{(\infty)}(t)$ are defined by
$$P^{t}_{n}(x) = \frac{\sin (n\theta)}{\sin \theta}, \,\,\, x=\sqrt{2\rho(t)}
\cos \theta\eqno(3.34)$$ 
and it is easy to check that
$$\int_{\Bbb R} P^{t}_{m}(x)P^{t}_{n}(x)\,d\mu_{t}(x)
= \delta_{mn},\eqno(3.35)$$
where
$$d\,\mu_{t}(x) = {\frac{\sqrt{2\rho(t) - x^2}}{\pi \rho(t)}}\,
1_{[\,-\sqrt{2\rho(t)}, \sqrt{2\rho(t)}\,\,]}(x)\,dx. \eqno(3.36)$$
Thus $d\mu_t$ is the spectral measure of $J^{(\infty)}(t)$ so that
$$(e_1, (J^{(\infty)}(t))^{k}\,e_1) =\int_{\Bbb R} x^{k} d\mu_{t}(x), \eqno(3.37)$$
Combining (3.32) and (3.37), we obtain the first part of the 
following theorem.

\proclaim
{Theorem 3.2.2} (a) For each $t>0,$ the sequence $(\mu^{n}_{t})_{n\geq 1}$ 
converges weakly, in probability, to the probability measure $\mu_{t}$
defined in (3.36).
\newline
(b) For each $t>0,$ the sequence $(\nu^{n}_{t})_{n\geq 1}$
converges weakly, in probability, to the same probability measure $\mu_{t}.$
\endproclaim

\demo
{Proof}  We have already proved (a).  In order to establish (b),
it suffices to show that
$$d_{LP}(\mu^{n}_{t},\nu^{n}_{t})\cp 0\quad \hbox{as}\,\,\, n\to\infty,\eqno(3.38)
$$
where $d_{LP}$ is the L\'evy-Prohorov metric on the space of (Borel) probability
measures on $\Bbb R.$  For this purpose, introduce the distribution functions
$F_{\mu^{n}_{t}},$ $F_{\nu^{n}_{t}}$ corresponding to $\mu^{n}_{t}$ and
$\nu^{n}_{t}$ respectively.  Then from the definitions of the L\'evy-Prohorov
metric and the L\'evy distance between distribution functions, we have
$$\eqalign{
d_{LP}(\mu^{n}_{t},\nu^{n}_{t}) & \leq \hbox{sup}\,\,|F_{\mu^{n}_{t}}(x)-
F_{\nu^{n}_{t}}(x)|\cr
& \leq \hbox{max}\,\,\left|\sum_{j=1}^{k} \mu^{(n)}_{j}(t) -{k\over n}\right|,
\cr}\eqno(3.39)$$
and so it suffices to show that
$$\hbox{max}\,\,\left|\sum_{j=1}^{k} \mu^{(n)}_{j}(t) -{k\over n}\right|\cp
0 \quad \hbox{as}\,\,\, n\to \infty.\eqno(3.40)$$
As $\sum_{j=1}^{k} \mu^{(n)}_{j}(t)$ has the Beta distribution (see (3.22)), 
the rest of the proof of identical to that of Theorem 5.4 in \c{BNR}.  For the
sake of completeness, we give the main steps.  First of all, from
the density of the Beta distribution, we have the moments of
$\sum_{j=1}^{k} \mu^{(n)}_{j}(t)$:
$$E\left[\,\left(\sum_{j=1}^{k} \mu^{(n)}_{j}(t)\right)^r\,\right]
=\frac{\Gamma\left({k\beta\over 2}+r\right)\Gamma\left({n\beta\over2}\right)}
{\Gamma\left({k\beta\over 2}\right)\Gamma\left({n\beta\over 2} +r\right)}
\eqno(3.41)$$
from which we see that
$$E\left[\sum_{j=1}^{k} \mu^{(n)}_{j}(t)\right]
= {k\over n}.\eqno(3.42)$$
By using (3.41) and its special case in (3.42), we find
$$E\left[\,\left|\sum_{j=1}^{k} \mu^{(n)}_{j}(t)-{k\over n}\right|^{4}\,\right]
= O\left({k(n-k)\over n^4}\right)\eqno(3.43)$$ so that
$$\sum_{k=1}^{n} E\left[\,\left|\sum_{j=1}^{k} \mu^{(n)}_{j}(t)-{k\over n}
\right|^{4}\,\right] = O\left({1\over n}\right).\eqno(3.44)$$
Hence for each $\epsilon >0,$ we obtain 
$$\eqalign{
P\left[\hbox{max}\,\,\left|\sum_{j=1}^{k} \mu^{(n)}_{j}(t) -{k\over n}\right|
> \epsilon \right] &
\leq\ \sum_{k=1}^{n} P\left[\,\,\left|\sum_{j=1}^{k} \mu^{(n)}_{j}(t) -
{k\over n}
\right| > \epsilon\,\, \right]\cr
& \leq\, {\epsilon^{-4}} \sum_{k=1}^{n} E\left[\,\left|\sum_{j=1}^{k} \mu^{(n)}_{j}
(t) -{k\over n}\right|^{4}\, \right]\cr
& = O\left({1\over {n\epsilon^{4}}}\right)\cr}\eqno(3.45)$$
from which (3.40) follows.

\pf
\enddemo

\bigskip
\bigskip
\subhead
4. \ The  beta-Laguerre and the beta-Wishart processes
\endsubhead

\bigskip

The matrix model of the $\beta$-Laguerre ensembles parametrized by
$a > -1$ \c{DE} is defined
schematically by
$$L_{\beta,a}\sim \pmatrix \frac{1}{\sqrt{\beta}} \chi_{(a+n)\beta}
 & \frac{1}{\sqrt{\beta}}\chi_{\beta(n-1)}  & 0 & 
\cdots\cr
0  & \frac{1}{\sqrt{\beta}}\chi_{(a+n-1)\beta} & 
 \frac{1}{\sqrt{\beta}}\chi_{(n-2)\beta} & \ddots\cr
\vdots & \ddots & \ddots & \ddots\cr
\endpmatrix ,\eqno(4.1)$$
where the entries on the diagonal and the subdiagonal of the
$n\times n$ matrix $L_{\beta,a}$ are independent.
Using the pdf of the Chi distribution, the joint density of the independent
entries of $L_{\beta,a}$ reads
$$W_{l}(x,y ) =  d_{n\beta} \prod_{i=1}^{n} x_{i}^{a+n-i+1)\beta -1} e^{-\frac{\beta}{2}
x_{i}^2}\prod_{i=1}^{n-1} y_{i}^{(n-i)\beta-1} e^{-\frac{\beta}{2} y_{i}^2},                
\eqno(4.2)$$
where $x_i = (L_{\beta,a})_{ii},$ $y_i = (L_{\beta,a})_{i,i+1}$ and
$$d_{n\beta} = \frac{2^{2n-1}(\beta/2)^{{na\beta\over 2}+{\beta\over 2} 
n^2}} {\prod_{j=1}^{n-1} \Gamma\left(j\beta\over 2\right) 
\prod_{j=1}^{n} \Gamma\left((a+j)\beta\over 2\right)}.\eqno(4.3)$$
In this section, we will present results analogous to those
in Section 3 for two related matrix models: the one defined in (4.1) and
an associated one consisting of matrices $L_{\beta,a}^{T}L_{\beta,a}.$
\medskip

\subhead
4.1 \ The beta Laguerre (Wishart) processes and the eigenvalue 
distribution
\endsubhead
\medskip

\definition
{Definition 4.1.1} The $\beta$-Laguerre process $(L_{\beta, a}(t))_{t\geq 0}$
parametrized by $a> -1$ is the stochastic process on $n\times n$
bidiagonal matrices $L_{\beta,a}(t)$ given by
$$L_{\beta, a}(t) = \pmatrix
\frac{1}{\sqrt{\beta}} R^{(a+n)\beta}(t) & \frac{1}{\sqrt{\beta}}
R^{(n-1)\beta}(t)  & 0 & \cdots\cr
0  & \frac{1}{\sqrt{\beta}} R^{(a+n-1)\beta}(t) & \frac{1}{\sqrt{\beta}}
R^{(n-2)\beta}(t) & \ddots\cr
\vdots & \ddots & \ddots & \ddots\cr
\endpmatrix\eqno(4.4)$$
where the processes on the diagonal and subdiagonal are
statistically independent of each other, and where $R^{\delta}(t)$ is the
generalized  Bessel process 
of dimension $\delta$ starting from $0$ with parameters $(1/2, 1).$
Let $J_{\beta, a}(t) = L_{\beta,a}(t)^{T} L_{\beta, a}(t),$  then the
process  $(J_{\beta, a}(t))_{t\geq 0}$ is called the associated
$\beta$-Wishart process.
\enddefinition

As in the analogous case in Proposition 3.1.2, the following is immediate 
from the product construction of independent Markov processes,
$P[L_{\beta,a}(0) =0] =1,$  (2.11) and (2.14).

\proclaim
{Proposition 4.1.2} The $\beta$-Laguerre process is the matrix-valued
diffusion process starting from $0$ with transition probability
density function 
$$P(t,\widetilde{L},L)= \beta^{n-\frac{1}{2}}
  \prod_{i=1}^{n} p^{(a+n-i+1)\beta}_{t}(\sqrt{\beta}\,\,\tilde{x}_i, 
  \sqrt{\beta}\,\,x_i)
  \prod_{j=1}^{n-1} p^{(n-i)\beta}_{t} (\sqrt{\beta}\,\,\tilde{y}_i,\sqrt{\beta}\,\,
   y_i),
\eqno(4.5)$$
with respect to Lebesgue measure $dxdy = dx_1\cdots dx_{n}dy_1\cdots dy_{n-1}$
on $\Bbb R_{+}^{n} \times \Bbb R_{+}^{n-1},$ where $\tilde{x}_i =\widetilde{L}_{ii}, 
\,\tilde{y}_{i}=\widetilde{L}_{i,i+1},$
$x_i = L_{ii}$ and $b_{i}= L_{i,i+1}.$  Thus the
joint density of the independent entries of $L_{\beta,a}(t)$ is given explicitly
by
$$P(t,0,L) = d_{n\beta}\, \rho(t)^{-{na\beta\over 2}-{\beta n^2\over 2}}
 \prod_{i=1}^{n} x_{i}^{(a+n-i+1)\beta -1} e^{-\frac{\beta x_i^2}{2\rho(t)}}
 \prod_{i=1}^{n-1} y_{i}^{(n-i)\beta-1} e^{-\frac{\beta y_i^2}{2\rho(t)}}.                
\eqno(4.6)$$
If $W_{l}(x,y)$ is the density in (4.1), we have
$$\lim_{t\to\infty} P(t,\widetilde{L}, L) = \int_{\Bbb{R}_{+}^n\times \Bbb{R}_{+}^n} 
W_{l}(\tilde{x},\tilde{y})P(t,\widetilde{L},L)\,\,d\tilde{x}d\tilde{y} = 
W_{l}(x,y).\eqno(4.7)$$ 

\endproclaim

We next calculate the joint probability density
function of the eigenvalues of 
$J_{\beta, a}(t) = L_{\beta,a}(t)^{T} L_{\beta, a}(t).$
To do that, we have to first compute the pushforward of the
measure $P(t,0,L)dxdy$ under the map $L\mapsto J=L^{T}L$
which sends $n\times n$ bidiagonal matrices to $n\times n$
Jacobi matrices. To this end, observe that
$$\eqalign{
P(t,0, L) = & \, \left(\frac{\beta}{\rho(t)}\right)^{n-\frac{1}{2}}
\prod_{i=1}^{n} p^{(a+n-i+1)\beta}_{\infty} \left(\frac{x_i}{\sqrt{\rho(t)}}\right)
\prod_{i=1}^{n-1} p^{(n-i)\beta}_{\infty} \left(\frac{y_i}{\sqrt{\rho(t)}}\right)\cr
= & \, \frac{1}{\rho(t)^{n-\frac{1}{2}}} W_{l} \left(\frac{x}{\sqrt{\rho(t)}}, 
      \frac{y}{\sqrt{\rho(t)}}\right).\cr}
\eqno(4.8)$$
Therefore, if we let $a_i = J_{ii} = (L^{T}L)_{ii},$ 
$b_i = J_{i,i+1} = (L^{T}L)_{i,i+1},$ then in the notations
of section 3 for Jacobi matrices, the calculations leading to the results
in the next proposition are similar to the corresponding
one in \c{DE} where one has to use (3.9), (3.15) and the fact
that the Jacobian of the map $L\mapsto J= L^{T}L$ is given by \c{DE}
$$2^{n} x_{n} \prod_{i=1}^{n-1} x_{i}^2.\eqno(4.9)$$

\proclaim
{Theorem 4.1.3} Under the $\beta$-Wishart process, the vector 
$\lambda(t)=(\lambda_1(t),\cdots,\lambda_n(t))$ of 
ordered eigenvalues of $J_{\beta,a}(t)$ and the vector 
$f(1,t) =(f_1(1,t),\cdots, f_n(1,t))$ of
first components of normalized eigenvectors are independent.
If $\Delta (\lambda)$ is the Vandermonde determinant,
then 
\smallskip
\noindent (a) the joint pdf of the unordered eigenvalues of 
$J_{\beta, a}(t)$ is given by
$$P_{n\beta}^{a}(t,\lambda)  = C_{n\beta}^{a}\, \rho(t)^{-{\beta a n\over 4}-
{\beta \over 4} n^2}
|\Delta(\lambda)|^{\beta} \prod_{i=1}^{n} \lambda_{i}^{{\beta\over 2}(a+1) -1}
\exp \left(-\frac{\beta}{2\rho(t)} \sum_{i=1}^{n}
\lambda_{i}^{2}\right),\eqno(4.10)$$
where
$$C_{n\beta}^{a} = \left(\beta\over 2\right)^{{\beta\over 4}a n +
{\beta\over 4} n^2} \prod_{j=1}^{n} \frac{\Gamma\left(1 +{\beta\over 2}\right)}
{\Gamma\left(1 + {j\beta\over 2}\right)\Gamma\left({(a+j)\beta\over 2}\right)}.\eqno(4.11)$$
\smallskip
\noindent (b) the joint density of $f_1(1,t),\cdots, f_n(1,t)$ with respect to
the measure $d\sigma$ on $S^{n-1}_{+}$ is given by
$$2^{n-1} \frac{\Gamma\left(n\beta\over 2\right)}{\left(\Gamma\left(\beta\over 2
\right)\right)^{n}} \prod_{i=1}^{n} q^{\beta -1}_{i}.\eqno(4.12)$$
\endproclaim

In view of (4.12), it follows that
if $\mu(t) = (\mu_1(t),\cdots, \mu_n(t))$ is the vector of
weights in the spectral measure of the spectral measure of
$J_{\beta,a}(t),$ then the distributions of 
$\mu_j(t)$ and $\sum_{j=1}^{k} \mu_j(t)$
are also given by (3.21) and (3.22) respectively.

\bigskip

\subhead
4.2 \ Time-dependent Marchenko-Pastur law and quarter-circle law
\endsubhead
\medskip

We introduce the following scaling of $L_{\beta, a}(t)$:
$$L^{(n)}_{\beta, a}(t) = \frac{L_{\beta, a}(t)}{\sqrt{n}},\eqno(4.13)$$
and let
$x^{(n)}_{k}(t) = (L^{(n)}_{\beta, a}(t))_{kk},$ 
$y^{(n)}_{k}(t) = (L^{(n)}_{\beta, a}(t))_{k,k+1}.$

We begin with the analog of Lemma 3.2.1 for the $\beta$-Laguerre
process.

\proclaim
{Lemma 4.2.1} As $n\to\infty,$
$$x^{(n)}_{k}(t)\cp \sqrt{\rho(t)},\quad y^{(n)}_{k}(t)\cp \sqrt{\rho(t)},
\eqno(4.14)$$
for each $t>0.$
\endproclaim

\demo
{Proof} For any $t>0$ and any $r=1,2,\cdots,$ it follows from 
the pdf of $R^{\delta}(t)$ that
$$\aligned
E[(y^{(n)}_{k}(t))^r] = & \frac{2\left({n\beta}\over {2 \rho(t)}\right)^{(n-k)\beta\over2}}
{\Gamma\left({(n-k)\beta}\over 2\right)}
\int_{0}^{\infty} x^{(n-k)\beta +r} \exp\left(-{\beta nx^2}\over 2 \rho(t)\right)
\,dx\\
= & \frac{2\left({2\rho(t)}\over {n\beta}\right)^{r\over2}}
{\Gamma\left({(n-k)\beta}\over 2\right)}
\int_{0}^{\infty} x^{(n-k)\beta -1 + r} e^{-x^2}\, dx\\
= & \frac{\left({2\rho(t)}\over {n\beta}\right)^{r\over2}}
{\Gamma\left({(n-k)\beta}\over 2\right)}\cdot
\Gamma\left({(n-k)\beta + r}\over 2\right)\\
\sim & \rho(t)^{r/2}\\
\endaligned
\eqno(4.15)
$$
as $n\to\infty$ where we have used the asymptotics of the Gamma
function.   Hence $y^{(n)}_{k}(t)\cp \sqrt{\rho(t)}$ as $n\to\infty.$
Similarly,
$$\aligned
E[(x^{(n)}_{k}(t))^r] = &\, \frac{2\left({n\beta}\over {2 \rho(t)}\right)^{(a+n-
k+1)\beta\over2}}
{\Gamma\left({(a+n-k+1)\beta}\over 2\right)}
\int_{0}^{\infty} x^{(a+n-k+1)\beta +r-1} \exp\left(-{\beta nx^2}\over 
2 \rho(t)\right)
\,dx\\
= & \, \left({2\rho(t)}\over {n\beta}\right)^{r\over2}\cdot \frac{
{\Gamma\left({(a+n-k+1)\beta + r}\over 2\right)} } {
\Gamma\left({(a+n-k+1)\beta}\over 2\right) }\\
\sim & \, \rho(t)^{r/2}\\
\endaligned
\eqno(4.16)
$$
as $n\to\infty$ and so we also have $x^{(n)}_{k}(t)\cp \sqrt{\rho(t)}.$
\pf
\enddemo

Now we introduce 
$$J^{(n)}_{\beta, a} (t) = (L^{(n)}_{\beta, a}(t))^{T} L^{(n)}_{\beta, a}(t)
\eqno(4.17)$$
and let $a^{(n)}_{k}(t) = (J^{(n)}_{\beta, a}(t))_{kk},$
$b^{(n)}_{k}(t) = (J^{(n)}_{\beta, a}(t))_{k,k+1}.$   Then from
the relations between the entries of $J^{(n)}_{\beta, a} (t)$
and $L^{(n)}_{\beta, a}(t),$ we can deduce the following 
from Proposition 4.2.1 when we invoke the continuous mapping
theorem.

\proclaim
{Corollary 4.2.2} As $n\to\infty,$
$$\aligned
& a^{(n)}_{1}(t)\cp \rho(t), \quad a^{(n)}_{k}(t)\cp 2\rho(t), \,k>1,\\
& b^{(n)}_{k}(t)\cp \rho(t), \,k\geq 1\\
\endaligned
\eqno(4.18)
$$
for each $t>0.$
\endproclaim

Let $\lambda^{(n)}_{1}(t),\cdots, \lambda^{(n)}_{n}(t)$ be the 
eigenvalues of $J^{(n)}_{\beta, a}(t).$   We consider the 
spectral measure process 
$$\mu^{n}_{t} = \sum_{j=1}^{n} \mu^{(n)}_{j}(t) \delta_{\lambda_{j}^{(n)}(t)}
\eqno(4.19)$$
and the empirical eigenvalue process
$$\nu^{n}_{t} = {1\over n} \sum_{i=1}^{n} \delta_{\lambda^{(n)}_i(t)}\eqno(4.20)$$
associated with $J^{(n)}_{\beta, a}(t).$   From the relations
$$\int_{\Bbb R} x^{k} d\mu^{n}_{t}(x) = (e_1, (J^{(n)}_{\beta, a}(t))^{k}\,e_1),\,\,
k=1,\cdots,
\eqno(4.21)$$
it follows that the moments $\int_{\Bbb R} x^{k} d\mu^{n}_{t}(x)$ are
polynomials in the entries of $J^{(n)}_{\beta, a}(t).$   Thus it follows
from Corollary 4.2.2 and the continuous mapping theorem that
$$\int_{\Bbb R} x^{k} d\mu^{n}_{t}(x)\cp (e_1, (J^{(\infty))}_{\beta, a}(t))^{k}\,e_1)
\eqno(4.22)$$
as $n\to\infty,$ where $J^{(\infty)}_{\beta, a}(t)$ is the Jacobi operator
on $\ell_{2}^{+}$ given by
$$J^{(\infty))}_{\beta,a}(t)= \pmatrix \rho(t) & \rho(t) & 0 & 0 & \cdots  \\
                   \rho(t) & 2\rho(t)  & \rho(t) & 0\cdots\\
                    0 & \rho(t) & 2\rho(t) & \rho(t) & 0 &\cdots \\
                   \cdots & \cdots & \cdots & \cdots &\cdots \endpmatrix. 
\eqno(4.23)$$
The next thing to do is to compute the spectral measure $d\mu_t$ of
$J^{(\infty)}_{\beta, a}(t).$  For this purpose, we will make use
of the method of Grosjean \c{G}.  First of all, we introduce the
(time-dependent) polynomials $\{P^{t}_{n}(x)\}_{n\geq 0}$ 
satisfying the recursion relations
$$P^{t}_{n+1}(x) = (x-2\rho(t)) P^{t}_{n}(x) - \rho(t)^{2} P^{t}_{n-1}(x),\,\,
n\geq 1\eqno(4.24)$$
and the initial conditions
$$P^{t}_{0}(x) =1, \quad P^{t}_{1}(x) = x -\rho(t).\eqno(4.25)$$
Also, introduce the functions of the second kind
$$Q^{t}_{n}(z) = \int_{\Bbb R} \frac{P^{t}_{n}(x)}{z-x}\, d\mu_{t}(x)
\eqno(4.26)$$
for $n\geq 0$ and for $z\in\Bbb C.$   From (4.24) and (4.25) above, we have
$$Q^{t}_{1}(t) = -1 + (z-\rho(t)) Q^{t}_{0}(z),\eqno(4.27)$$
and 
$$Q^{t}_{n+1}(z) = (z- 2\rho(t))Q^{t}_{n}(z) - \rho(t)^{2} Q^{t}_{n-1}(z),\,\,
n\geq 1,\eqno(4.28)$$
where $Q^{t}_{0}(z)$ is the Stieltjes transform of the spectral measure
$d\mu_{t}.$   By making use of these relations, we obtain the continued
fraction expansion
$$Q^{t}_{0}(z) = \frac{1} {z-\rho(t) -F^{t}(z)},\eqno(4.29)$$
where
$$F^{t}(z) = {\rho(t)^{2}\over\displaystyle z -2\rho(t)-
             {\strut \rho(t)^{2}\over\displaystyle z-2\rho(t)-
             {\strut \rho(t)^{2}\over\displaystyle z-2\rho(t)- \ldots}}}.
\eqno(4.30)
$$
But from the expression for $F^{t}(z),$ it is clear that
$$F^{t}(z) = \frac{\rho(t)^2}{z-2\rho(t)-F^{t}(z)}.\eqno(4.31)$$
Solving, we obtain
$$F^{t}(z) = \frac{z-2\rho(t) -\sqrt{(z-2\rho(t))^2-4\rho(t)^2}}{2},
\eqno(4.32)$$
where the branch of the square root is the one
which tends to the positive square root of $(x-2\rho(t))^2 - 4\rho(t)^2$
when $z$ tends to $x\in (4\rho(t), \infty).$   Therefore, when
we substitute (4.32) into (4.29), the result is
$$\int_{\Bbb R} \frac{d\mu_{t}(x)}{z-x} =\frac{1}
{{z\over 2} + {1\over 2}\sqrt{(z-2\rho(t))^2-4\rho(t)^2}}.\eqno(4.33)
$$
We next introduce the Fourier transform of $d\mu_t$:
$$\widehat\mu_{t}(x) = \int_{\Bbb R} e^{ixs}\,d\mu_{t}(s).\eqno(4.34)$$
Then from (4.33), its Laplace transform is given by
$$\eqalign{
{\Cal L}(\widehat\mu_{t})(p) & = \int_{0}^{\infty} e^{-xp} \widehat\mu_{t}(x)\,dx\cr
 & = -i \int_{\Bbb R} \frac{d\mu_{t}(s)}{-ip-s}\cr
 & = \frac{i}{{ip\over 2}-{1\over 2}\sqrt{(ip+2\rho(t))^{2} -4\rho(t)^{2}}}\cr}
\eqno(4.35)
$$
for $\hbox{Re}\,(p) >0.$
Thus by the inversion theorem for Laplace transform, we have
$${1\over 2\pi}\int_{\sigma-i\infty}^{\sigma+i\infty} \frac{e^{xp}}
  {{ip\over 2}-{1\over 2}\sqrt{(ip+2\rho(t))^{2} -4\rho(t)^{2}}}\,dp
  = \cases 0, & \text{$x > 0$}\\
           {1\over 2}\widehat \mu_{t}(0+), & \text{$x=0$}\\
           \widehat\mu_{t}(x), & \text{$x >0,$}\endcases
\eqno(4.36)
$$
where $\sigma$ is a real number which is greater than the real parts
of the singularities of ${\Cal L}(\widehat\mu_{t})(p)$ in the $p$-plane.
Now make the change of variable $p =iu$ in the integral on the left hand 
side of (4.36), this gives
$${-i\over 2\pi} \int_{-\infty -i\sigma}^{\infty-i\sigma} \frac{e^{ixu}}
{{u\over 2}+{1\over 2}\sqrt{(u-2\rho(t))^{2} -4\rho(t)^{2}}}
= \cases 0, & \text{$x > 0$}\\
           {1\over 2}\widehat \mu_{t}(0+), & \text{$x=0$}\\
           \widehat\mu_{t}(x), & \text{$x >0,$}\endcases
\eqno(4.37)
$$
where the path of integration is now a horizontal line below the 
singularities of the integrand in the $u$-plane.  But from
the above expression, it is easy to show that the denominator
of the integrand vanishes only at $u=0.$  Thus the 
set of singularities of the integrand coincides with the branch cut
$[0,4\rho(t)]$ of the function $\sqrt{(u-2\rho(t))^{2} -4\rho(t)^{2}}$
and hence we can take $\sigma = -\epsilon,$ where $\epsilon >0$ is
a small number.  But the fact that $d\mu_{t}$ is a real measure means
that $\overline{\widehat \mu_{t}(-x)} = \widehat\mu_{t}(x)$ for all $x.$
Consequently,
$$\widehat\mu_{t}(x)
 = {i\over 2\pi} \int_{\Bbb R} \left[\frac{e^{ixu}}
{{u\over 2}+{1\over 2}\sqrt{u(u-4\rho(t))}}\right]_{u=s-i\epsilon}^{u=s+i\epsilon}\,
ds, \eqno(4.38)
$$
for all $x\in \Bbb R$  where 
$\widehat\mu_{t}(0) ={1\over 2}(\widehat\mu_{t}(0+) + \widehat\mu_{t}(0-)).$
But from the definition of $\sqrt{u(u-4\rho(t))},$ we have
$$\lim_{\epsilon\to 0+} \sqrt{u(u-4\rho(t))}\mid_{u=s\pm i\epsilon} =
\cases -\sqrt{s(s-4\rho(t))}, & \text{$s\in (-\infty, 0]$}\\
       \pm i\sqrt{s(4\rho(t) -s)}, & \text{$s\in (0, 4\rho(t))$}\\
       \sqrt{s(s-4\rho(t))}, & \text{$s\in [4\rho(t), \infty).$}\endcases
\eqno(4.39)
$$
Therefore, when we take the limit as $\epsilon\to 0+$ in (4.38), the
only contribution to the integral comes from $[0, 4\rho(t)]$ and
we find
$$\widehat\mu_{t}(x) = \int_{0}^{4\rho(t)} e^{ixs} {1\over 2\pi\rho(t)}
\sqrt{\frac{4\rho(t)-s}{s}}\,ds.\eqno(4.40)$$
Hence we can now conclude that
$$d\mu_{t}(x) = {1\over 2\pi\rho(t)}
\sqrt{\frac{4\rho(t)-x}{x}}\,1_{(0,4\rho(t))}\, dx\eqno(4.41)$$
and so we have the following result.

\proclaim
{Theorem 4.2.3} (a) For each $t>0,$ the sequence $(\mu^{n}_{t})_{n\geq 1}$ 
converges weakly, in probability, to the probability measure $\mu_{t}$
defined in (4.41).
\newline
(b) For each $t>0,$ the sequence $(\nu^{n}_{t})_{n\geq 1}$
converges weakly, in probability, to the same probability measure $\mu_{t}.$
\endproclaim

We now return to the $\beta$-Laguerre process itself.  Note that
although we have Theorem 4.2.3 available to us, however, it is 
not hard to see that it is not possible to deduce 
the corresponding result for the $\beta$-Laguerre process.   
Our study of the $\beta$-Laguerre
process will be based on the following transformation which is well-known in 
numerical linear
algebra \c{GK}.  Suppose $B$ is the bidiagonal matrix
$$B = \pmatrix x_1 & y_1 & & &  \bigcirc\\
            & x_2 &  y_2  & &  \\
            & & \ddots & \ddots & \\
            & & &  x_{n-1} & y_{n-1}\\
           \bigcirc  & & & & x_n\endpmatrix
\eqno(4.42)
$$
with singular value decomposition $B = U\Sigma V^{T},$ where
$U = (u_1,\cdots, u_n)$ and $V = (v_1,\cdots, v_n)$ are
orthogonal, and $\Sigma =\hbox{diag}\,(\sigma_1,\cdots, \sigma_n)$ is the
diagonal matrix whose diagonal entries are the singular
values of $B.$  Then the eigenvalues of 
the $2n\times 2n$ symmetric tridiagonal matrix
$$S = \pmatrix 0 & x_1 &&&&& \bigcirc\\
               x_1 & 0 & y_1 &&&&\\
               & y_1& 0 &&&&\\
               & & & & \ddots &&\\
               && & \ddots &&& \\
               &&&  & & & x_n\\
               \bigcirc &&&& & x_n & 0\endpmatrix
\eqno(4.43)
$$
are $\sigma_1,\cdots, \sigma_n,-\sigma_1,\cdots, -\sigma_n$
and the first components of the corresponding normalized eigenvectors 
are given by ${v_1(1)/\sqrt{2}},\cdots, {v_n(1)/\sqrt{2}},$ $v_1(1)/\sqrt{2},$
$\cdots,$ $v_n(1)/\sqrt{2}$
respectively.  In view of this, it suffices to study the process
$(S_{\beta,a}(t))_{t\geq 0}$, where
$$S_{\beta,a}(t) = \pmatrix
0 & \frac{1}{\sqrt{\beta}}R^{(a+n)\beta}(t)  & 0 & 
\cdots\cr
\frac{1}{\sqrt{\beta}}R^{(a+n)\beta}(t) & 0 & 
 \frac{1}{\sqrt{\beta}}R^{(n-1)\beta}(t) & \ddots\cr
0  & \frac{1}{\sqrt{\beta}}R^{(n-1)\beta}(t) & 0& 
\ddots\cr
\vdots & \ddots & \ddots & \ddots\cr
\endpmatrix \eqno(4.44)$$
is obtained from (4.4) by applying the above transformation.
(The full justification is in Theorem 4.2.4 below.)
We introduce the following scaling of $S_{\beta,a}(t)$:
$$S^{(n)}_{\beta, a}(t) = \frac{S_{\beta, a}(t)}{\sqrt{n}},\eqno(4.45)$$
corresponding to (4.13).  By our discussion above,
if $\sigma^{(n)}_{1}(t),\cdots, \sigma^{(n)}_{n}(t)$ denote the
singular values of $L_{\beta, a}(t),$ then the eigenvalues of
$S^{(n)}_{\beta, a}(t)$ are given by
 $\sigma^{(n)}_{1}(t),$ $\cdots,$ $\sigma^{(n)}_{n}(t),$
$-\sigma^{(n)}_{1}(t),\cdots, -\sigma^{(n)}_{n}(t).$
Now we introduce the spectral measure process
$$\widetilde{\mu}^{2n}_{t} = \sum_{j=1}^{n} \widetilde{\mu}^{n}_{j}(t)
\left(\delta_{\sigma^{(n)}_j(t)} + \delta_{-\sigma^{(n)}_j(t)}\right)
\eqno(4.46)
$$
and the empirical eigenvalue
process 
$$\widetilde{\nu}^{2n}_t= {1\over {2n}} 
\sum_{i=1}^{n} \left(\delta_{\sigma^{(n)}_j(t)} + \delta_{-\sigma^{(n)}_j(t)}\right)
\eqno(4.47)
$$
associated with $(S^{(n)}_{\beta, a}(t))_{t\geq 0}$,
where
$$\widetilde{\mu}^{n}_j(t) = {1\over 2} f_j(1,t)^2,\quad j=1,\cdots,n.\eqno(4.48)$$


By using Lemma 4.2.1, and following the same procedure in
Section 3.2, we obtain
$$\int_{\Bbb R} x^{k} d\,\widetilde{\mu}^{2n}_{t}(x)\cp 
(e_1, (\sqrt{2}J^{(\infty))}(t))^{k}\,e_1)
= \int_{\Bbb R} x^{k} d\widetilde{\mu}_{t}(x),\eqno(4.49)$$
where $J^{(\infty)}$ is the Jacobi operator in (3.33) and 
$$ d\,\widetilde{\mu}_{t}(x)={\frac{\sqrt{4\rho(t) - x^2}}{2\pi \rho(t)}}\,
1_{[\,-2\sqrt{\rho(t)}, 2\sqrt{\rho(t)}\,\,]}(x)\,dx. \eqno(4.50)$$
Thus we have proved
$$d_{LP}(\widetilde{\mu}^{2n}_{t},\widetilde{\mu}_{t})\cp 0\quad 
\hbox{as}\,\,\, n\to\infty.\eqno(4.51)
$$
We next show that
$$d_{LP}(\widetilde{\mu}^{2n}_{t},\widetilde{\nu}^{2n}_{t})\cp 0\quad 
\hbox{as}\,\,\, n\to\infty.\eqno(4.52)
$$
Here the idea is also the same as before.  First we calculate
the distribution of the vector of weights
$$\widetilde{\mu}(t) = (\widetilde{\mu}^{n}_1(t),\cdots,\widetilde{\mu}^n_n(t))
\eqno(4.53)$$
which is a generalized Dirichlet distribution (we can think of the one defined
in (3.18) as the standard one) supported on the simplex
$$S(1/2)=\left\{(y_1,\cdots,y_{n-1})\Big| y_j\geq 0\,\, \hbox{for all}\,\,j, 
\sum_{j=1}^{n-1}y_j\leq 1/2\right\}.\eqno(4.54)$$ 
Then for each $1\leq k\leq n,$ we find
$$\sum_{j=1}^{k} \widetilde{\mu}^n_{j}(t) \sim \hbox{Beta}^{(0,\frac{1}{2})}
 \left({k\beta\over 2}, {(n-k)\beta\over 2}\right),\eqno (4.55)$$
 where $\hbox{Beta}^{(0,\frac{1}{2})}$ denotes the generalized Beta distribution
supported on $[0,\frac{1}{2}].$   Since we have
$$
d(\widetilde{\mu}^{2n}_{t},\widetilde{\nu}^{2n}_{t}) 
\leq \hbox{max}\,\,\left|\sum_{j=1}^{k} \widetilde{\mu}^{(n)}_{j}(t) -
{k\over 2n}\right|,\eqno(4.56)$$
the analysis proceeds as in Section 3.2.  Consequently, when we combine
(4.51) and (4.52), we conclude that
$$d(\widetilde{\nu}^{2n}_{t}, \widetilde{\mu}_{t})\cp 0\quad 
\hbox{as}\,\,\, n\to\infty.\eqno(4.57)
$$

We are now ready to
state the main result for the $\beta$-Laguerre process.
To this end, introduce the empirical singular value process
$$\bar{\nu}^{n}_t= {1\over {n}} 
\sum_{i=1}^{n} \delta_{\sigma^{(n)}_j(t)}
\eqno(4.58)
$$
associated with $(L^{(n)}_{\beta, a}(t))_{t\geq 0}.$
Also, let
$$d\,\bar{\mu}_t(x) = {\frac{\sqrt{4\rho(t) - x^2}}{\pi \rho(t)}}\,
1_{[\,0, 2\sqrt{\rho(t)}\,\,]}(x)\,dx. \eqno(4.59)$$
We will consider $\bar{\nu}^{n}_t$ and $\bar{\mu}_t$
as measures on $\Bbb R_{+} =[0, \infty).$

\proclaim
{Theorem 4.2.4} For each $t>0,$ the sequence $(\bar{\nu}^{n}_{t})_{n\geq 1}$
converges weakly, in probability, to the probability measure 
$\bar{\mu}_{t}.$
\endproclaim

\demo
{Proof} We will deduce this result from (4.57).  For this purpose,
it is more convenient to use the bounded Lipschitz metric, which is
equivalent to the L\'evy-Prohorov metric \c{Dud}. In order to 
write down the expression for this metric, denote by $BL(S)$ the 
class of bounded functions $f:S\longrightarrow \Bbb R$ on
a complete metric space $S$ which are Lipschitz.  For 
$f\in \, BL(S),$ define the Lipschitz semi-norm 
$$\|f\|_{L(S)} = \hbox{sup}_{x\neq y} \frac{|f(x)-f(y)|}{d(x,y)}\eqno(4.60)$$
and put $\|f\|_{BL(S)} = \|f\|_{L(S)} + \|f\|_{L^{\infty}(S)}$ where
$\|f\|_{L^{\infty}(S)}$ is the sup-norm.  Then $\|f\|_{BL(S)}$ is a norm
and $(BL(S), \|\cdot\|_{BL(S)})$ is a Banach space.  With these notations,
the bounded Lipschitz distance between the two measure $\widetilde{\nu}^{2n}_t$
and $\widetilde{\mu}_{t}$ is given by
$$d_{BL(\Bbb R)}(\widetilde{\nu}^{2n}_t, \widetilde{\mu}_t)=
  \hbox{sup}_{f\in BL_1(\Bbb R)} \Big|\int_{\Bbb R} f d\,\widetilde{\nu}^{2n}_t
  - \int_{\Bbb R} f d\,\widetilde{\mu}_t \Big|\eqno(4.61)$$
where the supremum is taken over
$$BL_1(\Bbb R) = \{f\in BL(\Bbb R)\mid \|f\|_{BL(\Bbb R)}\leq 1 \}.\eqno(4.62)$$
Now let $BL^{e}_1(\Bbb R) =\{f\in BL_1(\Bbb R)\mid f\, \hbox{is even}\},$
then clearly
$$d_{BL(\Bbb R)}(\widetilde{\nu}^{2n}_t, \widetilde{\mu}_t)\geq
  \hbox{sup}_{f\in BL^{e}_1(\Bbb R)} \Big|\int_{\Bbb R} f d\,
\widetilde{\nu}^{2n}_t - \int_{\Bbb R} f d\,\widetilde{\mu}_t \Big|.\eqno(4.63)
$$
But for $f\in BL^{e}_1(\Bbb R),$ it follows from the definition of the 
two measures that
$$\int_{\Bbb R} f d\,\widetilde{\nu}^{2n}_t = 
{1\over n} \sum_{j=1}^{n} f(\sigma^{(n)}_j(t)) 
=\int_{\Bbb R_{+}} fd\,\bar{\nu}^{n}_t,\eqno(4.64)
$$
while
$$\int_{\Bbb R} fd\,\widetilde{\mu}_t = \int_{\Bbb R_{+}} f(x) d\,\bar{\mu}_t.
\eqno(4.65)
$$
As we can identify the space  $BL^{e}_1(\Bbb R)$ with $BL_1(\Bbb R_{+}),$ 
when we combine (4.63)-(4.65), the result is
$$\eqalign{
d_{BL(\Bbb R)}(\widetilde{\nu}^{2n}_t, \widetilde{\mu}_t)\geq & \,
  \hbox{sup}_{f\in BL_1(\Bbb R_{+})} \Big|\int_{\Bbb R_{+}} f d\,
\bar{\nu}^{n}_t - \int_{\Bbb R_{+}} f d\,\bar{\mu}_t \Big|\cr
 = & \,d_{BL(\Bbb R_{+})} (\bar{\nu}^{n}_t, \bar{\mu}_t).\cr}\eqno(4.66)
$$
Hence it follows from (4.57) and (4.66) that
$d_{BL(\Bbb R_{+})}(\bar{\nu}^{n}_{t}, \bar{\mu}_{t})\cp 0\quad 
\hbox{as}\,\,\, n\to\infty.$

\pf
\enddemo

\remark
{Remark 4.2.5}  If we let 
$$\bar{\mu}^{n}_{t} = 2 \sum_{j=1}^{n} \widetilde{\mu}^{n}_{j}(t)
\delta_{\sigma^{(n)}_j(t)} =\sum_{j=1}^{n} f_j(1,t)^{2}\delta_{\sigma^{(n)}_j(t)} ,\eqno(4.67)$$ 
then following the same argument as 
in the proof of Theorem 4.2.4, we also have
$$d_{BL(\Bbb R_{+})}(\bar{\mu}^{n}_{t}, \bar{\mu}_{t})\cp 0\quad 
\hbox{as}\,\,\, n\to\infty.\eqno(4.68)$$
The measure valued process $(\bar{\mu}^n_t)_{t\geq 0},$ however, is
the spectral measure process of the square root 
$\left(\sqrt{J^{(n)}_{\beta,a}(t)}\,\right)_{t\geq 0}$ of the scaled
$\beta$-Wishart process, where $J^{(n)}_{\beta,a}(t)$ is defined
in (4.17).

\endremark

\enddocument